# RADIUS-Assisted Companion PSK Discovery: Enabling WPA3-SAE in Externally Authenticated Multi-PSK (ePSK / MPSK / DPSK / iPSK / PPSK) Deployments

**Azif Abdul Salam Kolisam beevi**[1,*], **Anil Kumar Reddy**[2], **Niraj Lochan Mahapatra**[3]

[1]Director of Enterprise Solutions Engineering, San Jose, CA, USA
[2]Principal Engineer, Bangalore, Karnataka, India
[3]Senior Technical Manager, Bangalore, Karnataka, India

[*]Corresponding author: azif.abdulsalam@gmail.com
Co-author emails: anilgeereddy@gmail.com, nirajlochan@gmail.com

## Abstract

Large multi-dwelling unit (MDU), student-housing and hospitality networks rarely store resident pre-shared keys (PSKs) on the access point (AP). Instead, the per-resident keys of an ePSK, MPSK, DPSK, iPSK or PPSK service are held in an external RADIUS server or a cloud database that is synchronised with the property-management system. In our earlier work, "Companion-BSSID Assisted PSK Discovery: Enabling WPA3-SAE in Multi-PSK (ePSK / MPSK / DPSK / iPSK) Community-Wide Wi-Fi Deployments" (CAPD), we showed that WPA3-SAE can be brought to multi-PSK networks by learning the client's PSK on a WPA2-only companion BSSID and steering the client with IEEE 802.11v to a hidden WPA2/WPA3 primary BSSID (the configured WLAN profile) of the same SSID. That scheme assumes the AP can test every PSK itself. This paper removes that assumption. We propose RADIUS-Assisted Companion PSK Discovery (RA-CAPD). The AP forwards the ANonce, the raw EAPOL-Key message 2 (SNonce, MIC, RSNE), the SSID, the client MAC, the BSSID and the AP MAC to the external server in RADIUS vendor-specific attributes (VSAs). The server performs MIC validation against all provisioned keys, identifies the matching PSK, and returns it to the AP encrypted under the RADIUS shared secret (or a per-AP key for cloud databases), together with the unit, room or user identity and the VLAN. The AP binds the client to this PSK and identity, synchronises the binding to every AP in the domain, and moves the client to the configured profile. 802.11v-capable clients are steered with a BSS Transition Management request; other clients are disconnected so that they reconnect to it. On the configured profile, SAE-capable clients complete WPA3-SAE and WPA2-only clients complete WPA2-PSK, both with the learned key and no further search. The companion BSSID is therefore used only for onboarding. The design, attribute dictionary, algorithms, complexity and security properties are presented, together with laboratory validation on commercial access points.

**Keywords:** WPA3, SAE, Multiple PSK, ePSK, DPSK, iPSK, MPSK, PPSK, RADIUS, Vendor-Specific Attribute, Cloud Authentication, MIC Validation, IEEE 802.11v, BSS Transition Management, Multi-Dwelling Unit, VLAN Isolation

## 1. Introduction

Managed Wi-Fi in multi-dwelling unit (MDU) properties, student housing, senior living and hotels is increasingly delivered as a single property-wide SSID on which every unit, room or resident is issued a unique pre-shared key (PSK) mapped to a private VLAN. The feature is marketed as Enterprise PSK (ePSK), Multiple PSK (MPSK), Dynamic PSK (DPSK), Identity PSK (iPSK) or Private PSK (PPSK); we refer to these collectively as *multi-PSK*. At small scale the PSK list can be configured on the AP or the WLAN controller. At property scale, however, keys are created, rotated and revoked by leasing workflows, so operators keep them in an **external authentication server**: an on-premises or hosted RADIUS server [1], or a cloud database belonging to the network-management or property-management platform.

Under WPA2-Personal the AP identifies the resident by testing the Message Integrity Code (MIC) of EAPOL-Key message 2 (M2) against every candidate PSK. Following the KRACK disclosures [2] and the offline-dictionary weakness of WPA2-PSK, WPA3-Personal replaces PSK-derived keys with Simultaneous Authentication of Equals (SAE) [3], [4], which by design allows only one password to be tested per exchange. MIC-style reverse lookup therefore no longer works, and the 6 GHz band admits only WPA3-SAE [5], [6]. Our earlier work [7] solved this with Companion-BSSID Assisted PSK Discovery (CAPD): a broadcast WPA2-only *companion* BSSID learns the client's PSK through MIC lookup, and IEEE 802.11v BSS Transition Management (BTM) [8] steers the client to a hidden WPA2/WPA3 *primary* BSSID of the same SSID, which carries the configured WLAN profile. There, SAE runs with the learned PSK.

CAPD assumes that the AP itself holds all N candidate PSKs. In externally authenticated deployments it does not, and should not: downloading thousands of resident keys to every AP enlarges the attack surface and creates a synchronisation problem. Today's RADIUS-based multi-PSK services work around this in one of two ways. Either the server looks up the client's MAC address and returns its PSK (MAC Authentication Bypass style), which fails with MAC address randomisation and requires pre-registration, or the server validates the MIC and returns only the VLAN, which is enough for WPA2 but leaves the AP unable to run SAE because it never learns the PSK.

This paper makes the following contributions:

1) We define the problem of WPA3 multi-PSK with an external key store and show why neither MAC-based PSK lookup nor VLAN-only MIC validation is sufficient to enable SAE.
2) We propose **RADIUS-Assisted Companion PSK Discovery (RA-CAPD)**. The AP exports the handshake material (ANonce, raw M2, SSID, client MAC, BSSID and AP MAC) to the RADIUS server or cloud database in vendor-specific attributes. The server performs the MIC validation and returns the matched passphrase, encrypted, together with the unit/room/user identity and VLAN.
3) We specify how the onboarding AP **synchronises the learned binding** (PSK, identity, VLAN, lifetime) to every AP in the domain, and how **every** onboarded client, whether SAE-capable or WPA2-only and whether 802.11v-capable or not, is moved to the configured profile, so that the companion BSSID is used only for onboarding.
4) We present the attribute dictionary, AP and server algorithms, computational and latency cost, and a security analysis, and we describe the laboratory validation on commercial access points.

Section 2 reviews background. Section 3 discusses related approaches. Section 4 presents the RA-CAPD architecture and attribute dictionary, and Section 5 the algorithms. Section 6 analyses complexity and latency, Section 7 security, Section 8 the experimental evaluation, and Section 9 limitations. Section 10 concludes.

## 2. Background

### *2.1 Multi-PSK with an External Key Store*

A multi-PSK service maintains tuples ⟨PSK_i, Identity_i, VLAN_i⟩, i = 1 … N, for a single SSID, where Identity_i is a unit number, room number or user name. Table 1 summarises where that database can live and what each location implies for the AP.

**Table 1. Multi-PSK key-store deployment models**

| Key store | Who validates the MIC | Implication for WPA3 (SAE) |
|---|---|---|
| AP-local / controller | AP or controller | AP knows PSK_k; CAPD [7] applies directly. |
| RADIUS server (on-prem or hosted) | RADIUS server (this paper) | AP must *receive* PSK_k from the server before it can run SAE. |
| Cloud database / cloud RADIUS | Cloud service (this paper) | Same as RADIUS; transport may be RADIUS, RadSec or an HTTPS API. |
| MAC-registered PSK (MAB style) | None (MAC lookup) | Breaks with randomised MACs; needs pre-registration. |

Whatever the vendor term (ePSK, MPSK, DPSK, iPSK or PPSK), the key the client uses determines the identity and therefore the VLAN. Each VLAN is a separate layer-2 broadcast domain, so one resident's devices cannot discover or reach another's, while every resident can roam across the whole property on one SSID.

### *2.2 WPA2-PSK Four-Way Handshake and MIC Validation*

In WPA2-Personal the Pairwise Master Key is derived from the passphrase and SSID [9]:

$$PMK = PBKDF2\text{-}HMAC\text{-}SHA1(PSK, SSID, 4096, 256) \quad (1)$$

The AP (authenticator address AA, here the BSSID) sends ANonce in M1; the client (supplicant address SPA) returns SNonce and a MIC in M2:

$$PTK = PRF\text{-}384(PMK, \text{"Pairwise key expansion"}, min(AA,SPA) \parallel max(AA,SPA) \parallel min(ANonce,SNonce) \parallel max(ANonce,SNonce)) \quad (2)$$

$$MIC = HMAC\text{-}SHA1\text{-}128(KCK, \text{EAPOL-Key M2 with MIC field zeroed}), \quad KCK = PTK[0..127] \quad (3)$$

Every input except the PMK is visible in M1 and M2. Any party that holds the candidate PMKs and receives **ANonce, M2, SSID, AA and SPA** can therefore find the unique k for which Eqs. (2)–(3) reproduce the received MIC. RA-CAPD relies on this: the set of values the AP must export to the server is exactly the set of free variables in Eqs. (1)–(3). The AP MAC is added for scoping and audit.

### *2.3 SAE, the WPA3 Gap and the CAPD Approach*

SAE [3], [4] maps the password to a group element PWE (or, with hash-to-element, to a password element PT) and runs a balanced PAKE in which each exchange tests exactly one password. An AP must commit to the right password *before* the exchange, so it cannot search N keys. CAPD [7] separates key *discovery* (WPA2 on the companion BSSID) from key *use* (the configured profile on the primary BSSID). Both BSSIDs share the SSID string, so the client keeps one network profile and, under per-SSID MAC randomisation, presents the same MAC address on both. In RA-CAPD, discovery moves to the external server, while key use stays on the AP. **The AP therefore needs the PSK itself, not just the VLAN.**

### *2.4 RADIUS Attribute Model and Key Protection*

A RADIUS attribute is encoded as Type, Length and Value [1]. Vendor-defined information is carried in Vendor-Specific Attributes (VSAs, Type 26), which add a 4-octet Vendor-Id (an IANA Private Enterprise Number) followed by a Vendor-Type, a Vendor-Length and the value. Keys returned in Access-Accept are traditionally protected with the salt-based MD5 key-hiding scheme of RFC 2548 [10], keyed by the RADIUS shared secret and the Request Authenticator of the corresponding Access-Request. Attributes of the same type are kept in order end to end, and proxies must not reorder them [1], [11].

### *2.5 IEEE 802.11v BSS Transition Management*

IEEE 802.11v [8], [3] BTM lets an AP recommend a target BSSID to a client that advertises BSS Transition support (Extended Capabilities bit 19). As in CAPD [7], RA-CAPD uses BTM to steer the client from the companion BSSID to the primary BSSID once the PSK is known. Clients without BTM support are disconnected instead (Section 5.5).

## 3. Related Work and Alternative Approaches

Table 2 compares the approaches available for per-resident segmentation with an external key store against the requirements of an MDU deployment.

1) **AP / controller-held PSK database (CAPD [7]).** This works with WPA3 but requires every AP or controller to hold all N keys. It does not apply when the operator's key store is a RADIUS server or a cloud database, and it duplicates sensitive keys across the network.
2) **MAC-based PSK lookup (MAB-style iPSK).** The AP sends the client MAC to RADIUS and receives the PSK for that MAC. This supports SAE, but each device MAC must be pre-registered, and it fails under MAC randomisation, which is now on by default in major mobile and desktop operating systems.
3) **Server-side MIC validation returning VLAN only.** The server validates the M2 MIC and returns the VLAN, and in some designs the PMK. This is sufficient for WPA2, but the AP never learns the PMK, so it cannot derive the SAE password element, and the client cannot be moved to WPA3 or 6 GHz.
4) **SAE password identifiers.** The client names its password in SAE Commit [3], and the AP can then query RADIUS for that identifier. This is standards-native but requires client-side identifier configuration, which mainstream operating systems do not expose.
5) **WPA3-Enterprise (802.1X).** This gives per-user credentials and RADIUS VLANs [12], [13], but many residential devices (TVs, consoles, IoT) cannot run 802.1X, and resident onboarding is complex.

**Table 2. Comparison of approaches for WPA3 multi-PSK with an external key store**

| Approach | Keys stay on server | WPA3 / 6 GHz | No client config change | Randomised-MAC safe | Works with IoT |
|---|---|---|---|---|---|
| AP / controller DB (CAPD) | No | Yes | Yes | Yes | Yes |
| MAC-based PSK lookup | Yes | Yes | Yes | No | Yes |
| MIC validation, VLAN only | Yes | No | Yes | Yes | Yes |
| SAE password identifier | Yes | Yes | No | Yes | No |
| WPA3-Enterprise | Yes | Yes | No | Yes | No |
| **RA-CAPD (proposed)** | **Yes**[1] | **Yes** | **Yes** | **Yes** | **Yes**[2] |

[1] APs hold only the PSKs of bound clients (Section 5.4), never the full database. [2] WPA2-only devices use WPA2-PSK on the configured profile with the learned key (Section 5.5).

To the best of our knowledge, this is the first work to combine server-side MIC validation, encrypted PSK return, domain-wide binding synchronisation and 802.11v steering to deliver WPA3 multi-PSK from an external key store with unmodified clients.

## 4. Proposed Architecture

RA-CAPD keeps the over-the-air behaviour of CAPD unchanged and adds a RADIUS (or cloud API) exchange between M2 and M3 of the companion-BSSID handshake. No client modification is required. Figure 1 shows the architecture.

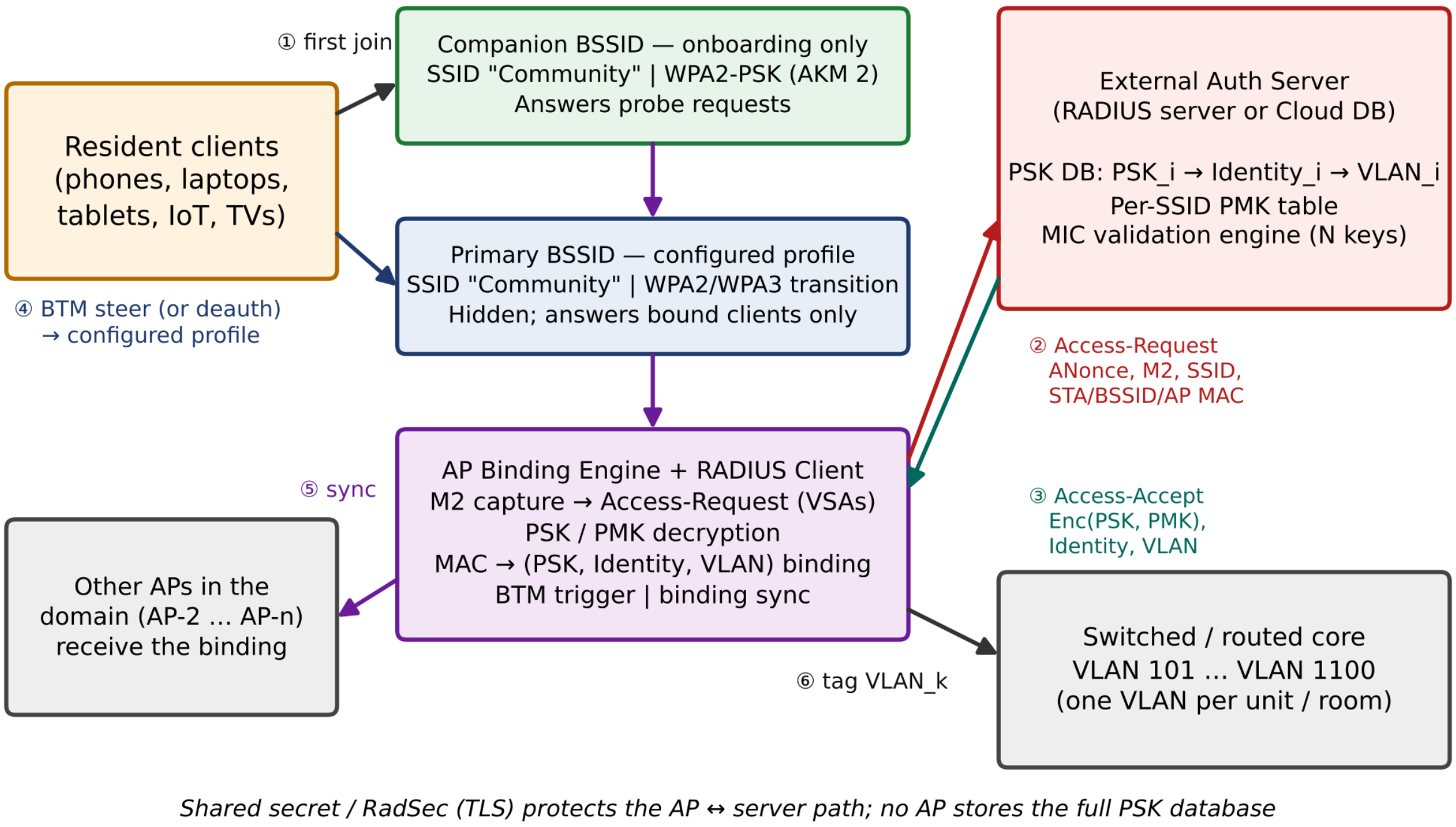

**Figure 1.** RA-CAPD architecture: the AP exports handshake material to the external server, which validates the MIC and returns the encrypted PSK, identity and VLAN; the AP synchronises the binding to the other APs in the domain and moves the client to the configured profile.

### *4.1 Components*

- **Companion BSSID (C-BSS).** Advertises the community SSID with WPA2-PSK (AKM 2) only and is used **only for onboarding**. It answers probe requests from unknown stations, holds **no** PSK list, and on receiving M2 hands the handshake to the RADIUS client and holds M3. For a short steering window after onboarding it ignores probe requests and denies association from bound MACs, so that they reconnect to the configured profile.
- **Primary BSSID (P-BSS), the configured profile.** Uses the same SSID string in WPA2/WPA3 transition mode (AKM 2 + AKM 8) on 2.4/5 GHz and WPA3-SAE-H2E only on 6 GHz. It is hidden and answers only stations with a valid binding. It runs SAE, or WPA2-PSK for WPA2-only clients, using the PSK held in the binding. It never performs a PSK search.
- **AP RADIUS client and Binding Engine.** Builds the Access-Request with the VSAs of Table 4, verifies the Access-Accept, decrypts the PSK and PMK, and stores the binding B[MAC_STA] = ⟨PSK_k, Identity_k, VLAN_k, expiry⟩.
- **Binding Synchronisation.** Distributes B[MAC_STA] from the onboarding AP to every AP in the domain (Section 5.4), so that any AP can serve the client on the configured profile.
- **External Authentication Server (EAS).** A RADIUS server or cloud database holding ⟨PSK_i, Identity_i, VLAN_i⟩ and a pre-computed per-SSID PMK table. It performs MIC validation (Algorithm 2), records the proven binding MAC → Identity_k, and returns the encrypted keys.

- **Steering Agent.** Issues the 802.11v BTM Request to the primary BSSID immediately after M4 to BTM-capable clients, and Disassociate the others.

## *4.2 WLAN Profile Configuration*

**Table 3. Recommended WLAN profile configuration**

| Parameter | Companion BSSID (onboarding) | Primary BSSID (configured profile) |
|---|---|---|
| SSID string | Community (same) | Community (same) |
| Security (2.4 / 5 GHz) | WPA2-PSK (AKM 2) only | WPA2/WPA3 transition (AKM 2 + 8), PMF capable |
| Security (6 GHz) | Not configured | WPA3-SAE, H2E only (AKM 8/24), PMF required |
| PSK source | External server (MIC validation) | Synchronised binding table |
| SSID broadcast / probe | Broadcast; answers unbound STAs | Hidden; answers bound STAs only |
| 802.11v BTM | Enabled (sender) | Enabled |
| Clients served | Onboarding only | All onboarded clients (SAE and WPA2-only) |
| VLAN assignment | Tunnel-Private-Group-ID from Accept | VLAN_k from binding |

## *4.3 RADIUS Attribute Dictionary*

Table 4 lists the attributes the AP sends in the Access-Request, and Table 5 the attributes the server returns. Vendor attributes use the operator's or AP vendor's Private Enterprise Number (PEN); the Vendor-Type numbers below are illustrative. Standard attributes are used wherever their semantics fit [1], [13], so the request remains readable by ordinary RADIUS policy engines.

**Table 4. Access-Request attributes sent by the AP (companion BSSID, after M2)**

| Attribute | Type | Length (octets) | Content / purpose |
|---|---|---|---|
| User-Name (1) | Std | 12–17 | Client MAC, used as session identity |
| Calling-Station-Id (31) | Std | 17 | Client MAC (SPA in Eq. 2) |
| Called-Station-Id (30) | Std | ≤ 50 | "BSSID:SSID" per RFC 3580 |
| NAS-Identifier (32) / NAS-IP-Address (4) | Std | var. / 4 | AP identity |
| NAS-Port-Type (61) | Std | 4 | 19 = Wireless-IEEE-802.11 |
| Message-Authenticator (80) | Std | 16 | HMAC-MD5 integrity; mandatory [14], [15] |
| MPSK-ANonce (VSA 1) | VSA | 32 | ANonce generated by the AP in M1 |
| MPSK-EAPOL-M2 (VSA 2) | VSA | ~121 | Raw EAPOL-Key M2 as received: SNonce, MIC, RSNE / Key Data |
| MPSK-SSID (VSA 3) | VSA | 1–32 | SSID octets where the client associates (PMK salt) |
| MPSK-Client-MAC (VSA 4) | VSA | 6 | Client MAC in binary (SPA) |
| MPSK-BSSID (VSA 5) | VSA | 6 | BSSID MAC (AA in Eq. 2) |
| MPSK-AP-MAC (VSA 6) | VSA | 6 | AP base / Ethernet MAC: site scoping, audit |

**Table 5. Access-Accept attributes returned by the server**

| Attribute | Type | Length (octets) | Content / purpose |
|---|---|---|---|
| MPSK-Encrypted-Keys (VSA 10) | VSA | 114 | Salt + encrypted {PSK_k, PMK_k, lifetime} (Eq. 4); fits in one VSA |
| MPSK-Identity (VSA 11) | VSA | ≤ 64 | Unit / room number or user name linked to PSK_k, e.g. "Bldg-B/Unit-1204" |
| Tunnel-Type (64) / Tunnel-Medium-Type (65) | Std | 6 / 6 | VLAN (13) / IEEE-802 (6) [16], [13] |
| Tunnel-Private-Group-ID (81) | Std | var. | VLAN_k |
| Session-Timeout (27) / MPSK-Binding-TTL (VSA 12) | Std / VSA | 4 | Binding lifetime on the APs |
| Message-Authenticator (80) | Std | 16 | Integrity of the whole Accept |

The encrypted payload consists of the passphrase (≤ 63 octets), the 32-octet PMK and a 4-octet lifetime, each encoded as a 2-octet-header TLV (105 octets). With the 2-octet salt and padding to 16-octet blocks it occupies 114 octets and is carried in a single VSA. If no PSK matches, the server returns Access-Reject and the AP deauthenticates the client with reason code 15 (4-way handshake timeout), exactly as a local MIC failure would.

## 5. The RA-CAPD Algorithm

RA-CAPD has three phases: WPA2 onboarding on the companion BSSID (Phase 1), external MIC validation and key return (Phase 1b), and operation on the configured profile (Phase 2). Figure 2 shows the message exchange, Figure 3 the AP-side decision logic, Figure 4 the server-side logic and Figure 5 the binding synchronisation.

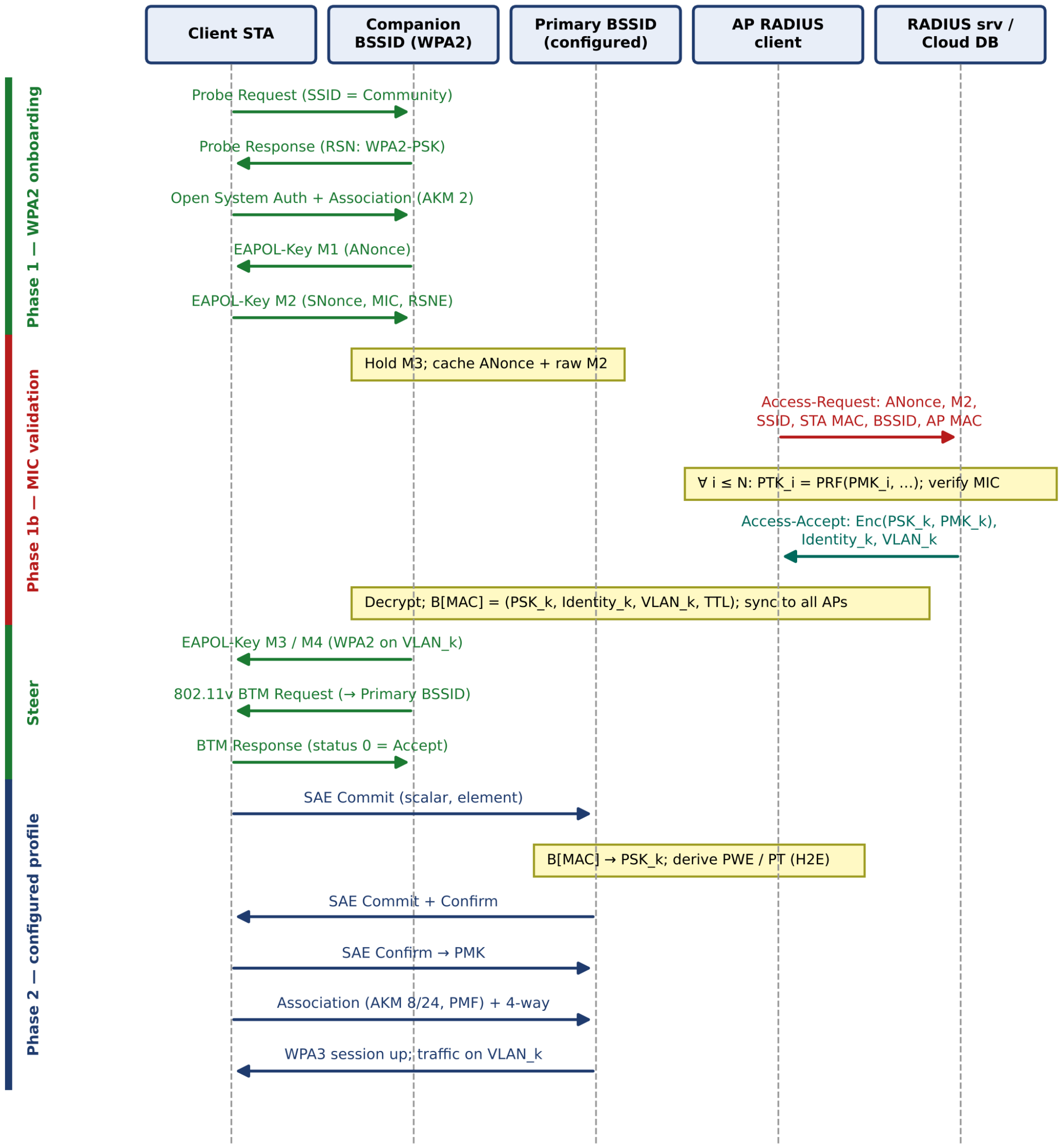

**Figure 2.** RA-CAPD packet exchange: WPA2 onboarding, RADIUS/cloud MIC validation with encrypted PSK return, binding sync, 802.11v steering, and WPA3-SAE association on the configured profile with the learned PSK.

### *5.1 Phase 1: WPA2 Onboarding on the Companion BSSID*

1) A new client finds only the broadcast companion BSSID (the primary BSSID is hidden and does not answer unknown stations), authenticates with Open System and associates with AKM 2.
2) The AP sends EAPOL-Key M1 with a fresh ANonce and keeps a copy of the ANonce.
3) The client derives the PTK from the passphrase configured for the "Community" profile and returns M2 with SNonce, MIC and RSNE.
4) The AP stores the raw M2 octets and **holds M3**, because it does not yet know which PMK to use.

### *5.2 Phase 1b: External MIC Validation*

1) The AP sends an Access-Request carrying the attributes of Table 4: **ANonce**, the **raw M2** (SNonce, MIC, etc.), the **SSID**, the **client MAC**, the **BSSID** and the **AP MAC**, protected by Message-Authenticator. For a cloud database the same fields are sent as a structured message over a mutually authenticated TLS connection.

2) The server selects the candidate PMK table for the SSID, optionally narrowed to the site or AP group identified by the AP MAC. For each candidate it derives $PTK_i$ (Eq. 2) and recomputes the MIC over M2 with the MIC field zeroed (Eq. 3). It stops at the first match k. With high-entropy keys a collision is cryptographically negligible.
3) On a match, the server records the proven binding $MAC_{STA} \rightarrow Identity_k$ with a timestamp. It then encrypts the actual $PSK_k$, together with $PMK_k$ so that the AP can send M3 without running PBKDF2, under the RADIUS shared secret (Eq. 4) or a per-AP key, and returns an Access-Accept with the encrypted keys, **MPSK-Identity** (unit, room or user), VLAN attributes and binding TTL.

### *5.3 Key Return and Client Tagging*

The AP verifies the Response Authenticator and Message-Authenticator and decrypts the keys. The RFC 2548-style key hiding [10] for plaintext $P = p_1 \parallel p_2 \parallel \ldots$ (16-octet blocks) is:

$$b_1 = MD5(S \parallel RA \parallel Salt),\ \ c_1 = p_1 \oplus b_1;\ \ \ b_i = MD5(S \parallel c_{(i-1)}),\ \ c_i = p_i \oplus b_i \qquad (4)$$

where S is the RADIUS shared secret, RA the Request Authenticator of the Access-Request and Salt a 2-octet value with its most significant bit set, unique per attribute. Because RA is fresh for each request, a captured Accept cannot be replayed against a different handshake. When the transport is RadSec [17] or DTLS [18], or a cloud API, the keys are instead sealed with AES-256-GCM [19] under a per-AP key. The AP then:

- completes M3/M4 with $PMK_k$ and places the client in $VLAN_k$;
- stores $B[MAC_{STA}] = \langle PSK_k, Identity_k, VLAN_k, now + TTL \rangle$, so the client is now **tagged with its actual PSK and its unit/room/user identity**, which appear in AP logs, client tables and accounting records;
- zeroes the received ciphertext and keeps $PSK_k$ only in protected memory for the lifetime of the binding.

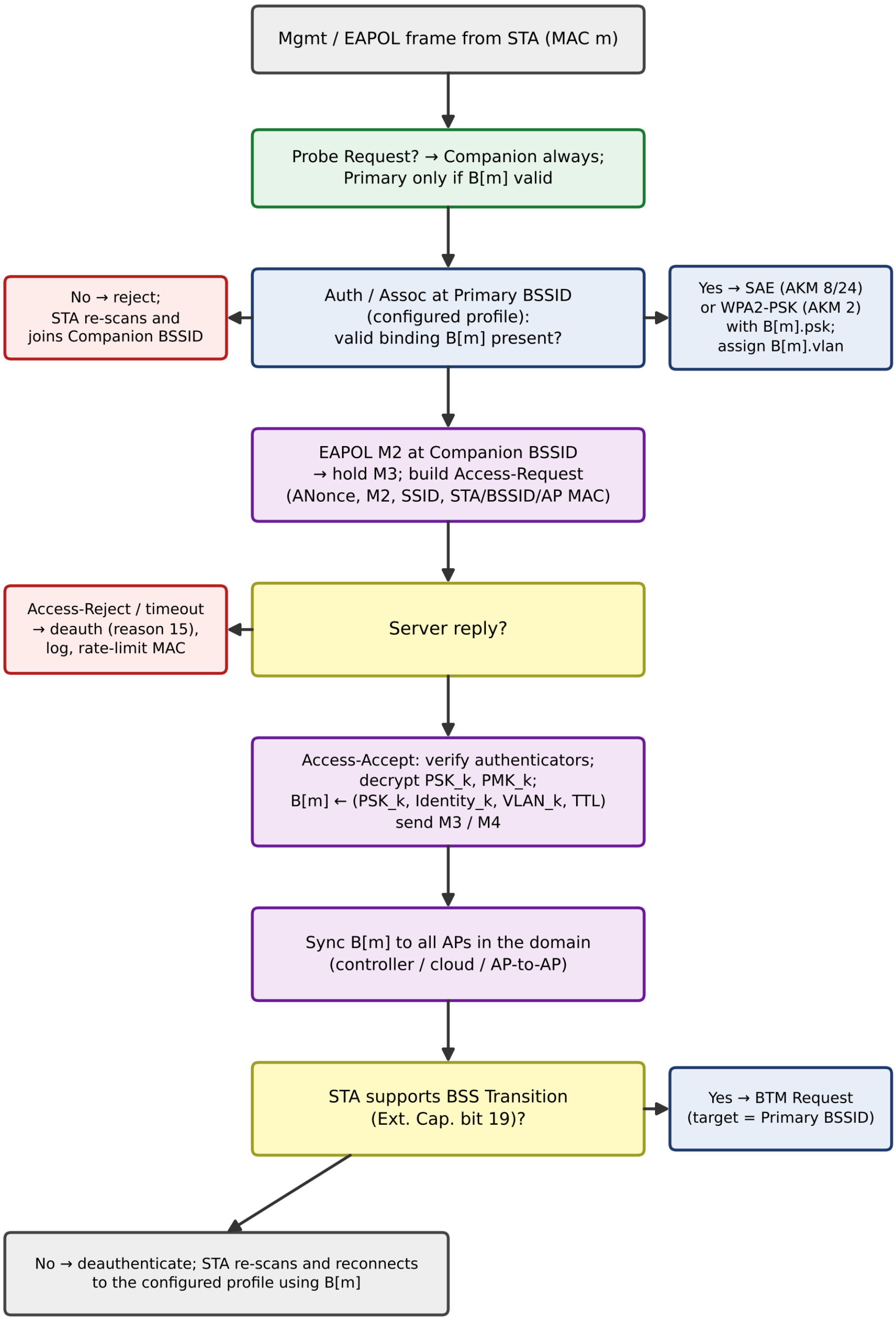

**Figure 3.** AP-side decision flow for RA-CAPD.

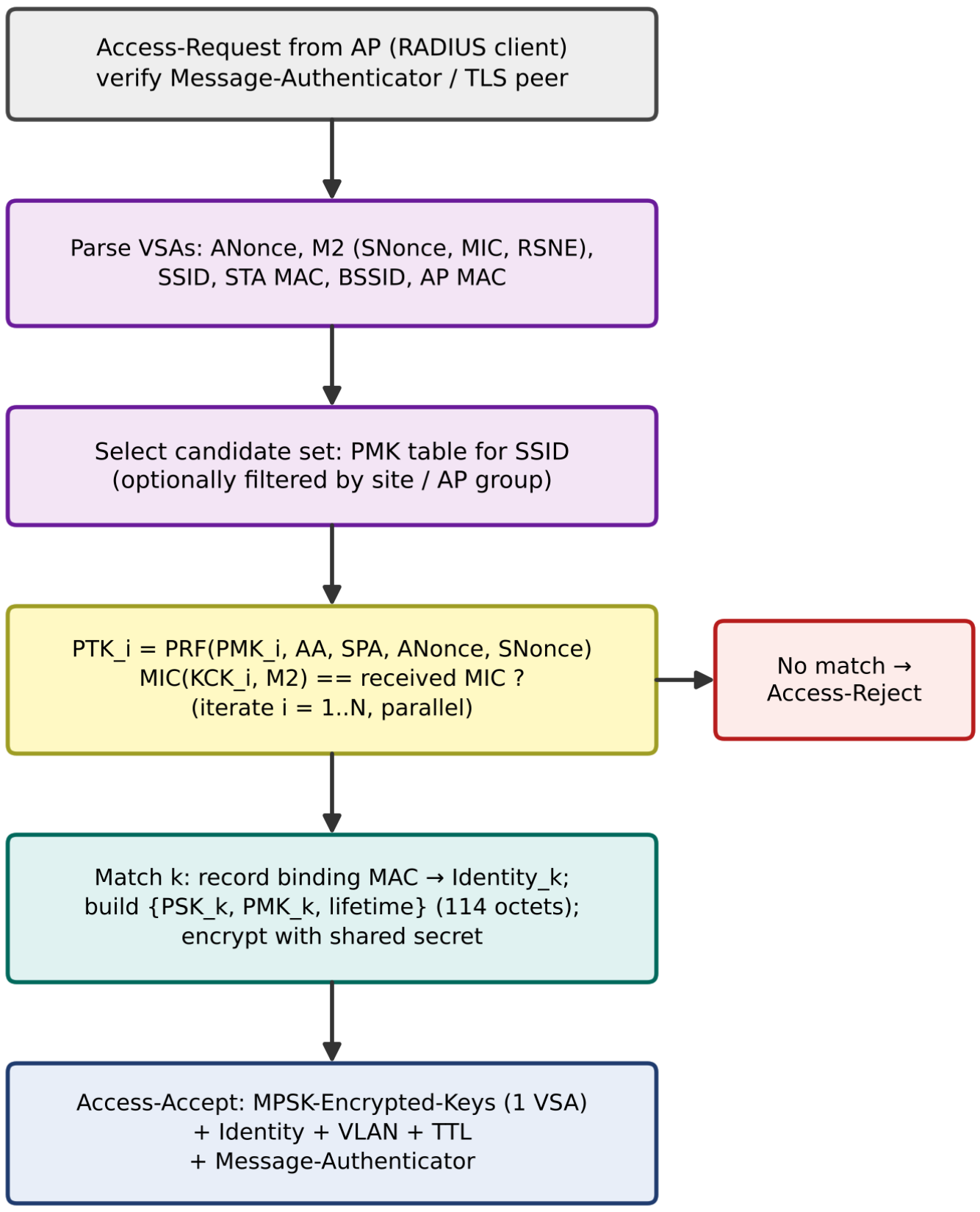

**Figure 4.** Server-side (RADIUS / cloud DB) MIC validation and key-return flow.

### *5.4 Binding Synchronisation Across the Domain*

The AP that onboarded the client is the only AP that talks to the external server for that client. As soon as it has stored B[MAC_STA], it **synchronises the binding**, comprising the client MAC, the encrypted PSK_k, Identity_k, VLAN_k and the remaining lifetime, to **all other APs in the domain** (the site, AP group or roaming domain that serves the SSID). Synchronisation runs over the existing management plane: through the WLAN controller, through the cloud management service, or by authenticated AP-to-AP messaging. The PSK is always encrypted in transit under a management-plane key. Figure 5 shows the exchange.

After synchronisation every AP holds the binding, so a client that roams, reconnects after a disconnection, or lands on a different AP after steering is served immediately on the configured profile, with **no new WPA2 onboarding and no further RADIUS exchange**. When the operator revokes or rotates a PSK, the server sends a Disconnect-Request or CoA [20] to the onboarding AP, which propagates a Binding Delete to all APs. The client is deauthenticated and, if it still holds a valid key, re-onboards. Bindings that reach their TTL are removed on every AP.

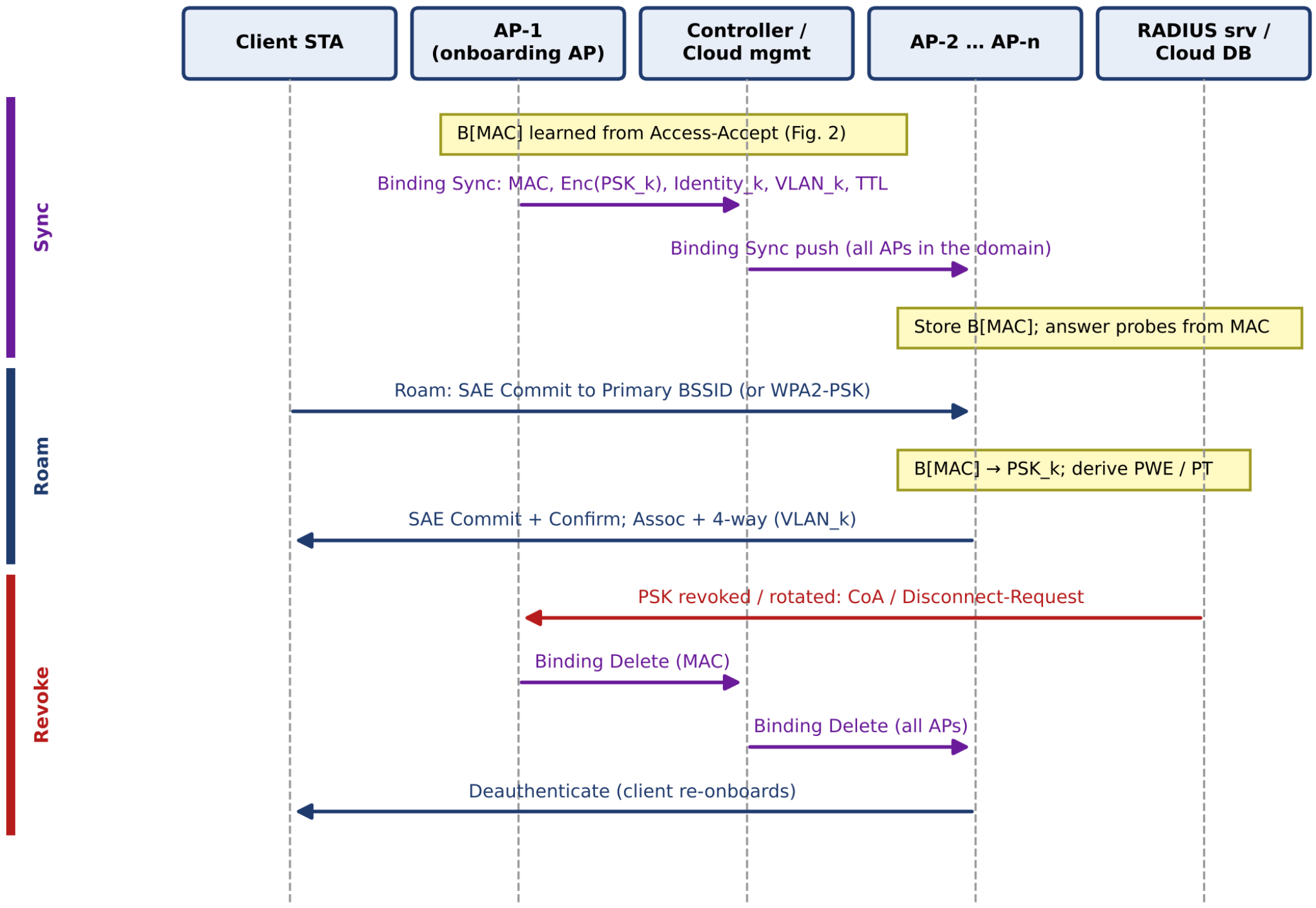

**Figure 5.** Binding synchronisation to all APs in the domain, roaming on the configured profile without RADIUS, and revocation.

## 5.5 Phase 2: Moving Every Client to the Configured Profile

The companion BSSID exists only to onboard clients. Once the PSK is known, **every** client is moved to the configured profile on the primary BSSID, whatever its capabilities:

- **802.11v-capable clients** receive a BTM Request nominating the primary BSSID (the 6 GHz primary first for tri-band SAE clients) and reassociate.
- **Clients without 802.11v** are disassociate once the binding is stored. When they re-scan, the companion BSSID does not answer their MAC during the steering window, and the primary BSSID answers it from the binding, so they reconnect to the configured profile.
- **SAE-capable clients** start SAE with the same passphrase. The AP finds B[MAC_STA], derives PWE/PT from PSK_k, and SAE succeeds on the first attempt. The client associates with AKM 8/24, PMF is negotiated, and traffic is tagged with VLAN_k.
- **WPA2-only clients** (for example older IoT devices) select AKM 2 on the transition-mode configured profile. The AP uses PMK_k from the binding directly, so the four-way handshake completes **without any PSK search**, and the client receives VLAN_k with the same isolation as on the companion BSSID.

If a client arrives at the primary BSSID without a valid binding (for example after MAC rotation or binding expiry), the AP rejects it. The client re-scans, finds the companion BSSID and repeats onboarding transparently.

## 5.6 Algorithm Specification

**Algorithm 1. RA-CAPD, AP side**

```
Input : SSID S; companion BSSID C (WPA2-PSK, AKM 2, onboarding only)
        primary BSSID P (configured profile: WPA2/WPA3, hidden)
        EAS = RADIUS server or cloud DB; shared secret / per-AP key K
State : binding table B : MAC -> (psk, pmk, identity, vlan, expiry)
```

```
on EAPOL_M2(m, frame) at C:
    hold M3
    req <- AccessRequest(User-Name = m, Calling-Station-Id = m,
             Called-Station-Id = C.bssid ":" S,
             VSA ANonce, VSA EAPOL-M2 = frame, VSA SSID = S,
             VSA Client-MAC = m, VSA BSSID = C.bssid, VSA AP-MAC,
             Message-Authenticator)
    resp <- send(EAS, req, timeout = T_rad, retries = 1)
    if resp is Reject or timeout:
        deauth(m, reason = 15); rate_limit(m); return
    (psk, pmk) <- Decrypt(resp.EncryptedKeys, K, req.Authenticator)
    send M3 using pmk; await M4; assign resp.vlan to m
    B[m] <- (psk, pmk, resp.Identity, resp.vlan, now + resp.TTL)
    SyncBinding(B[m], all APs in domain)       // Section 5.4
    if BTM_capable(m):
        send BTM_Request(m, candidates = [P_6GHz, P_5GHz, P_2.4GHz])
    else:
        deauth(m)                              // STA reconnects to P
    C.suppress(m, T_steer)                     // C ignores m for a while

on Auth/Assoc(m, akm) at P:
    if B[m] invalid: reject(m); return     // client falls back to C
    if akm in {8, 24}: run SAE with PWE/PT <- derive(B[m].psk, S)
    else:              run 4-way handshake with B[m].pmk   // AKM 2
    associate(m, akm, PMF as negotiated); assign B[m].vlan
```

**Algorithm 2. RA-CAPD, server side (RADIUS server or cloud DB)**

```
Input : DB D = {(PSK_i, ID_i, VLAN_i)}
        per-SSID table PMK[S][i] = PBKDF2(PSK_i, S, 4096, 256)

on AccessRequest(req) from AP:
    verify Message-Authenticator (or TLS peer certificate)
    (A, M2, S, m, aa, apmac) <- parse VSAs
    (SN, MIC) <- parse(M2)
    for i in candidates(S, site(apmac)):     // parallel over cores
        PTK <- PRF384(PMK[S][i], aa, m, A, SN)
        if HMAC_SHA1_128(PTK.KCK, zero_mic(M2)) == MIC:
            bind(m -> ID_i, now)
            E <- Encrypt(TLV(PSK_i) | TLV(PMK[S][i]) | TLV(TTL),
                         secret, req.Authenticator)   // Eq. (4) / GCM
            return AccessAccept(MPSK-Encrypted-Keys = E,
                      MPSK-Identity = ID_i, VLAN_i, TTL,
                      Message-Authenticator)
    return AccessReject
```

### *5.7 Cloud Database Variant*

When the key store is a cloud database reached through an HTTPS or WebSocket API, the AP sends the same six fields (ANonce, M2, SSID, client MAC, BSSID, AP MAC) as a structured message over mutually authenticated TLS. The cloud returns the PSK and PMK sealed with AES-256-GCM under a per-AP key established when the AP is onboarded to the cloud. The cloud management service that hosts the key store is also a natural place to run the binding synchronisation of Section 5.4.

## 6. Computational Complexity and Latency

Let N be the number of PSKs provisioned for the SSID. PBKDF2 (Eq. 1) depends only on the PSK and SSID, so the server pre-computes the PMK table once per key and SSID. Per onboarding the server performs at most N evaluations of PRF-384 and HMAC-SHA1-128, N/2 on average. This is the same O(N) cost as AP-side lookup [7], now executed on server-class CPUs and trivially parallel across cores; for property-scale N the search can additionally be narrowed with the AP-MAC/site attribute.

The end-to-end delay inserted between M2 and M3 is:

$$T_hold = RTT(AP \leftrightarrow EAS) + T_search(N) + T_decrypt \quad (5)$$

For an on-premises RADIUS server, the RTT is typically a few milliseconds. For a cloud database it is typically tens to low hundreds of milliseconds. T_hold should remain well below the client's handshake timers. We recommend an AP RADIUS timeout T_rad of 1 s with one retry, and that the AP retransmit M1 with the *same* ANonce if T_hold approaches the supplicant's retry interval. The server can then still validate the original M2. Phase 2 adds no RADIUS traffic, because every AP uses the synchronised binding.

**Table 6. Per-client cost summary**

| Operation | When | Cost |
|---|---|---|
| PMK pre-computation | Key provisioning / rotation | N × PBKDF2 per SSID (server, offline) |
| MIC validation | Phase 1b: first join, MAC change, binding expiry | ≤ N × (PRF + HMAC) on server; 1 RADIUS round trip |
| Key decryption | Every Accept | 7 MD5 blocks or 1 AES-GCM; O(1) |
| Binding synchronisation | Once per onboarding | 1 management-plane message per AP (or one multicast) |
| BTM Request / Response or deauth | Phase 1 → 2 | 2 management frames (or 1 Disassociate + reconnect) |
| Binding lookup | Every association on P-BSS | O(1), local |
| SAE or WPA2-PSK handshake | Every association / roam | 1 standard exchange; no RADIUS, no search |

## 7. Security Analysis

RA-CAPD inherits the client-side properties of CAPD [7]: a standard WPA3-SAE session with PMF in steady state, including the post-Dragonblood implementation hardening [21], and WPA2 exposure limited to the onboarding handshake. This section focuses on what is new, namely the AP ↔ server exchange, the key return and the binding synchronisation. Table 7 summarises the analysis.

### *7.1 Confidentiality of the Returned PSK*

The Access-Accept carries real key material. With classic RADIUS/UDP it is protected only by the MD5-based hiding of Eq. (4), whose strength rests on the entropy of the shared secret. Operators should therefore (i) use long random shared secrets (≥ 32 characters), (ii) prefer RadSec (RADIUS/TLS) [17] or RADIUS/DTLS [18], with AES-256-GCM sealing of the keys under a per-AP key, and (iii) never carry the RADIUS/UDP leg across untrusted networks. The ANonce, SNonce and MIC sent in the request are already visible over the air and reveal nothing new.

### *7.2 Integrity and Blast-RADIUS*

The Blast-RADIUS attack [15] showed that an on-path attacker can forge an Access-Accept from an Access-Reject when Message-Authenticator is absent. RA-CAPD therefore **requires** Message-Authenticator in every request and response and requires the AP to reject any response without it, consistent with current guidance [14]. Because the keys are encrypted under values bound to the Request Authenticator, a forged Accept would in any case fail decryption.

### *7.3 Key Exposure on the APs*

APs receive the PSKs of clients that have proven possession of them with a valid MIC, and through synchronisation every AP in the domain holds those bound keys. A compromised AP therefore exposes the keys of clients bound in its domain, but never the database or the keys of residents who have not onboarded, which is a strict improvement over AP-held key lists. Bindings are synchronised encrypted, kept in protected memory, and erased on expiry, revocation (CoA / Disconnect [20]) or reboot.

### *7.4 The Binding Is Not a Credential*

As in CAPD, spoofing a bound MAC address gains an attacker nothing. The binding only tells the AP which PSK to use, and SAE or the four-way handshake fails without the PSK. The binding and its PSK are delivered only to APs, never to stations.

### *7.5 Onboarding Exposure and Downgrade*

Phase 1 is a WPA2-PSK handshake, so a passive attacker who captures it could attempt an offline dictionary attack against that resident's passphrase. This exposure is identical to, and less frequent than, that of today's WPA2 multi-PSK networks, and high-entropy generated PSKs make it impractical. As in any transition-mode network, an attacker could try to keep an SAE-capable client on WPA2; monitoring the proportion of SAE-capable clients that associate with AKM 2 detects this.

**Table 7. Threat and mitigation summary**

| Threat | Impact under RA-CAPD | Mitigation |
|---|---|---|
| Eavesdropping on AP ↔ server | PSK exposure if secret weak | RadSec / DTLS; AES-GCM; strong secret |
| Forged / modified Access-Accept | Wrong PSK or VLAN | Mandatory Message-Authenticator; RA-bound encryption |
| AP compromise | Keys of bound clients only | No database on APs; TTL; CoA revocation |
| Eavesdropping on binding sync | PSK exposure | Encrypted, authenticated management plane |
| MAC spoofing of binding | Handshake fails without PSK_k | Binding is not a credential |
| Offline dictionary (onboarding) | Phase-1 capture only | High-entropy PSKs; phase 1 infrequent |
| Downgrade to WPA2 | Same as transition mode | Monitoring; periodic BTM; band restriction |

## 8. Experimental Evaluation

### *8.1 Testbed*

RA-CAPD was implemented in the firmware of an enterprise-class tri-band access point and evaluated in a laboratory environment representative of an MDU deployment. The testbed comprised 4 Cambium X7-35X APs managed by cnMaestro X, an external RADIUS server extended with the MIC-validation module of Algorithm 2, and a cloud database endpoint. The external server was provisioned with 1,000 resident PSKs, each mapped to a unit identity and a unique VLAN. The APs held **no** PSKs. Over-the-air traffic was captured with an additional AP in sniffer mode, and RADIUS traffic was captured on the AP uplink to verify attribute contents.

### *8.2 Methodology*

Each client was provisioned with its resident's passphrase for the "Community" network and run through the following test cases:

1) **TC1 External validation:** the client joins the companion BSSID; the server identifies the correct PSK among all candidates from the Access-Request VSAs.
2) **TC2 Key return and tagging:** the Access-Accept carries the encrypted PSK and unit/room identity; the AP decrypts it, completes M3/M4, and shows the client tagged with its identity and VLAN.

3) **TC3 Binding synchronisation:** the binding appears on all APs in the domain within the synchronisation interval.
4) **TC4 BTM steering and SAE:** a BTM-capable client is steered to the configured profile and completes SAE with AKM 8/24 and PMF.
5) **TC5 Non-11v client:** the client is deauthenticated after onboarding and reconnects to the configured profile.
6) **TC6 WPA2-only client:** the client reconnects to the configured profile with AKM 2 using the learned key.
7) **TC7 VLAN isolation and roaming:** ARP, ICMP and mDNS/SSDP discovery towards other residents' devices fail; the client roams between APs on the configured profile without RADIUS traffic.
8) **TC8 Failure handling:** wrong passphrase (Access-Reject), server unreachable, and PSK revocation (CoA) each lead to a clean deauthentication.

### *8.3 Results*

The encrypted keys (114 octets) fit in a single VSA, and onboarding, steering and SAE behaved as in the AP-local CAPD design [7]. Table 8 records the interoperability results by client category.

**Table 8. Interoperability results by client category**

| Client category | Example device / OS | TC1 | TC2 | TC3 | TC4 | TC5 | TC6 | TC7 | TC8 |
|---|---|---|---|---|---|---|---|---|---|
| iOS / iPadOS | iPhone 17 Pro Max (iOS 26.6.2) | PASS | PASS | PASS | PASS | N/A | N/A | PASS | PASS |
| Android | Pixel 6 Pro (Android 15) | PASS | PASS | PASS | PASS | N/A | N/A | PASS | PASS |
| macOS | MacBook Pro (macOS 26.4) | PASS | PASS | PASS | PASS | N/A | N/A | PASS | PASS |
| IoT (WPA2-only) | Ring Video Doorbell 2 | PASS | PASS | PASS | N/A | PASS | PASS | PASS | PASS |
| Windows 11 | Windows 11 laptop | PASS | PASS | PASS | PASS | PASS | N/A | PASS | PASS |

N/A marks test cases that do not apply to a device class (for example TC5 applies only to clients without 802.11v). Highlighted cells are to be completed with the measured results; adjust the N/A marks to the measured 802.11v support of each device.

## 9. Discussion and Limitations

- **Dependence on the server path.** First-time onboarding requires the server to be reachable. Already-bound clients continue to reconnect and roam from the synchronised binding during a server outage. Operators should deploy redundant servers and size T_rad accordingly.
- **Synchronisation scope and delay.** A client that reaches a second AP before the binding has arrived is rejected once and re-onboards. The synchronisation domain should match the roaming domain of the SSID, and the synchronisation delay should be small relative to typical reconnection times.
- **Key material in transit.** Returning the actual PSK to the AP is unavoidable if the AP is to run SAE. Its protection depends on the transport and the shared secret, which is why RadSec / DTLS and AES-GCM sealing are strongly recommended (Section 7.1).
- **Vendor attributes.** The VSAs require dictionary entries on the AP and the server. Standard attributes are reused wherever possible.

- **Search cost at very large N.** Properties with tens of thousands of keys on a single SSID should scope candidates by site or building and parallelise the search. The cost per onboarding remains linear and occurs only at first join, MAC rotation or binding expiry.
- **Standards path.** Native SAE password identifiers [3] with RADIUS lookup would make server-side MIC validation unnecessary. RA-CAPD delivers WPA3 multi-PSK with an external key store using unmodified clients today.

## 10. Conclusion and Future Work

Externally authenticated multi-PSK services keep resident keys in a RADIUS server or cloud database, so the AP cannot perform the MIC reverse lookup on which multi-PSK identification and our earlier CAPD scheme depend. This paper presented RADIUS-Assisted Companion PSK Discovery. The AP exports the ANonce, raw EAPOL-Key M2, SSID, client MAC, BSSID and AP MAC in RADIUS VSAs. The server validates the MIC against every provisioned key and returns the matched PSK, encrypted under the shared secret, together with the unit, room or user identity and VLAN. The AP binds and tags the client, synchronises the binding to every AP in the domain, and moves every client to the configured profile: by IEEE 802.11v steering where supported and by disconnection otherwise. SAE-capable clients then run WPA3-SAE with the learned key, and WPA2-only clients run WPA2-PSK without a search. The scheme keeps the key database off the APs, needs no client modification, preserves per-resident VLAN isolation, and extends server-managed multi-PSK to WPA3 and 6 GHz.

Future work includes standardisation of the attribute dictionary, hardware-accelerated MIC search for very large key stores, large-scale field measurements of T_hold over cloud paths, and extension of the binding to Wi-Fi 7 Multi-Link Operation.

## Acknowledgment

The authors thank Cambium Networks for the laboratory facilities used to test and validate this work.